# Tailoring Properties of Magneto-Optical Photonic Crystals

Amir Djalalian-Assl
51 Golf View Drive,
Craigieburn, VIC 3064,
Australia
amir.djalalian@gmail.com

**Abstract**: Magneto-Optic Photonic Crystals (MOPC) used in low dimension lasers whereby acting as Faraday rotators capable of 45º rotation where insertion loos is compensated by photoluminescence effect are of significant interest. In low dimension sensors MOPCs act as optical filters to selectively detect only lights at certain wavelengths. Combined with the photoluminescence effect, sensors may be designed capable of detecting signals with extremely low intensity, thus high quantum efficiency. In this work, MOPC with various Erbium dopant concentrations, acting as photoluminescence center, were modelled under high pumping power regime associated to 0.985 inversion population to identify the minimum Erbium concentration and number of layers for the target 45º FR, 0.9 transmittance and minimal ellipticity at the resonance wavelength 1531nm. Further optimization of Magneto-optical (MO) properties versus the microcavity position within the MOPC was investigated. An optimum multilayered configuration and composition was identified with a reduction in both erbium concentration and total thickness compared to what was reported previously without compromising the target MO properties.

# 1. Introduction

Bismuth-doped rare-earth garnet crystals exhibit high Faraday Rotation (FR) and are utilized for non-reciprocal passive optical devices in telecom applications. These garnet materials being highly transparent at near-infrared telecommunication wavelengths, have applications in sensors, polarization dependent and independent isolators, circulators, switches, and interleavers. Specific FR grows linearly with the Bi-content reaching -3.1 deg/µm for $\lambda$ = 633 nm in $Bi_{1.5}Y_{1.5}Fe_5O_{12}$ which has the highest Bi concentration possible in iron garnets. Completely substituted $Bi_3Fe_5O_{12}$ (BIG) films have been sintered by reactive ion beam sputtering [1] and pulsed laser deposition (PLD) techniques [2]. BIG is a record holder for FR effect among the magnetic garnets showing specific FR of $\theta_F = -8.4º$ /µm at 633 nm [1]. At longer wavelengths, FR gradually falls down whereas film becomes more transparent. Therefore, for MO-applications there is a tradeoff between FR and absorption to achieve superior MO-figure of merit represented by $Q[\deg] = \theta_F \times$ film thickness/ln(1/transmittance). Although BIG has a record FR $\theta_F = -0.35º$ /µm at $\lambda$ = 1.5 µm, (i.e. *C*-band), to get 45º FR rotation, BIG crystal should be as thick as 130 µm.

One dimensional Magneto-optical photonic crystals (MOPCs) might be considered as perspective technical solution to fabricate 45º thin film Faraday rotators. MOPCs utilize the idea of localizing the light inside a microcavity having a thickness $d = p\lambda_s/2n_{BIG}(\lambda_s)$, where *p* is an integer greater than zero, sandwiched between two Bragg mirrors, designed for the resonance wavelength $\lambda_s$. Bragg mirror contains several reflector pairs each having a thickness $d = (2q+1)\lambda_s/4n_{BIG}(\lambda_s)$, where *q* is an integer equal to or greater than zero. Due to the non-reciprocity, Faraday Effect has a multiplicative property in garnets: i.e. FR increases by a factor of $N$, where $N$ is number of times light experiences reflections between the Bragg mirrors. Presently, all-garnet heteroepitaxial $[Bi_3Fe_5O_{12}/Sm_3Ga_5O_{12}(SGG)]^m$ MOPCs grown by rf-magnetron sputtering demonstrate the best MO performance achieved so far. E.g. $[BIG/SGG]^5 BIG^2 [SGG/BIG]^5$ MOPC designed for $\lambda_s = 980$ nm showed FR $= -8.4^o$ and $Q = 43.6^o$ that was 630% enhancement compared to a single layer BIG film [3]. To reach $45^o$ FR at $\lambda = 980$ nm the total thickness of MOPC must be increased. This can be achieved by various methods by increasing the number of reflectors in two Bragg mirrors or, as it was proposed by Levy et al [4], using multiple resonant microcavities. Any method chosen to increase the total FR, deals with increasing the total thickness of magneto-optical Bi-contained material, as a result, drastically losing MOPC transmittance at $\lambda_s$. Luminescent MOPCs may present a solution to the high insertion loss in MOPCs.

Intensive luminescence is a much needed condition to produce optically gain media. Using $Er^{3+}$-ion as a luminescent agent enables realization of new type of gain garnet media. Since $Er^{3+}$-ion easily substitutes any rare earth element occupying the dodecahedral sites in the garnet structure therefore it can be added as dopant to both the garnet layers in Bragg reflectors and/or microcavities. Recently, room temperature photoluminescence (PL) was observed in several rare earth gallium and iron contained garnets [5]. Pumping garnet films with Ar-laser at 514.5 nm, a strong PL was observed at $\lambda = 980$ nm in Er-doped $La_3Ga_5O_{12}$ and $Gd_3Ga_5O_{12}$ garnets whereas there was no noticeable PL at 980 nm in iron contained Er-doped $Y_3Fe_5O_{12}$ and $Bi_3Fe_5O_{12}$ garnets. On the contrary, at $\lambda = 1531$ nm (*C*-band), $Er:Y_3Fe_5O_{12}$ and $Er:Bi_3Fe_5O_{12}$ show five times stronger PL than gallium contained garnets. Although Er-doped garnets exhibit PL properties, the giant FR present in $Bi_3Fe_5O_{12}$ is not affected noticeably. $Bi_{2.9}Er_{0.1}Fe_5O_{12}$ film shows specific FR with a peak value of -30 deg/µm at 535 nm and equals -6.7 deg/µm and -1.63 deg/µm at 654 nm and 980 nm, respectively [6].

The main idea proposed here is a new technical solution to design an Amplifying Magneto-Optical Photonic Crystal (AMOPC) from Er-doped garnets, capable of $45^o$ or more FR having a transmittance not less than 0.7. AMOPCs are anticipated to compensate for high insertion loss present in undoped MOPCs.

In order to investigate the feasibility of the proposed AMOPC, theoretical analysis and modelling techniques based on various microscopic approaches, namely Swanepoel formula and Višňovský's 4×4 matrix formalism was deployed and a simulation program was developed in MATLAB.

# 2. Theoretical Background

## 2.1 Properties of Light

Optical properties of light such as Polarization, FR and Ellipticity are of particular interests in applications of MOPCs since they are transformed as light propagates through MOPC. In this section, a brief overview of optical properties of light, extracted from [7], will be presented to the reader.

Recounting the interrelation between different states of polarizations, I shall only resort to basic optics where neither Stocks parameters nor Jones vector would be mention since, although they provide a convenient notation for the states of polarization and their superposition, they neither explain the physics behind those phenomena nor do they play any role in modelling techniques to come.

A linearly polarized light, P-states of polarization, is constituted from two orthogonal electric field vectors, $E_x$ and $E_y$, and a relative phase difference of $\theta_{phase} = 0$ or integral multiples of $\pm 2\pi$. Accordingly, mathematical representation of a linearly polarized electromagnetic wave having its electric field vectors in *x-y* plane which is propagating in z direction towards an observer takes the form:

$$\mathbf{E_x} = \hat{\imath}\, E_{0x} cos(kz - \omega t)$$

$$\mathbf{E_y} = \hat{\jmath}\, E_{0y}\, cos(kz - \omega t)$$

Whose vector sum leads to:

$$\mathbf{E} = [\hat{\imath}\, E_{0x} + \hat{\jmath}\, E_{0y}]\, cos(kz - \omega t)$$

It is apparent that the resultant electric field vector has a fixed amplitude and direction. Note that a phase difference of odd integer multiple of $\pm\pi$ would also result in a linearly polarized light whose polarization plane has rotated compared to the previous scenario.

A circularly polarized light, i.e. *R*- or *L*-states of polarizations, can be constituted from two orthogonal electric components, $E_x$ and $E_y$, having equal amplitude, (i.e. $E_{0x}=E_{0y}=E_0$) and a relative phase difference of $\theta_{phase} = -\pi/2$. Mathematical representation in this case takes the form:

$$\mathbf{E_x} = \hat{\imath}\, E_0 cos(kz - \omega t)$$

$$\mathbf{E_y} = \hat{\jmath}\, E_0\, cos\left(kz - \omega t - \frac{\pi}{2}\right) = \hat{\jmath}\, E_0\, sin(kz - \omega t)$$

$$\rightarrow \mathbf{E} = E_0[\hat{\imath}\, cos(kz - \omega t) + \hat{\jmath}\, sin(kz - \omega t)]$$

The resultant wave in this case has an electric field component whose direction is time varying. The resultant electric vector is rotating clockwise, i.e. Right Circularly Polarized (RCP), at an angular frequency, $\omega$, considering light moving towards an observer looking at $z$-direction. In case of a Left Circularly Polarized (LCP) light, $\theta_{\text{phase}} = \pi/2$ which leads to the resultant electric field vector:

$$\mathbf{E} = E_0[\hat{\imath}\cos(kz - \omega t) - \hat{\jmath}\sin(kz - \omega t)]$$

It is possible to synthesize a linearly polarized light from superposition of two oppositely circularly polarized lights, e.g. LCP and RCP, leading to resultant electric field vector:

$$\mathbf{E} = \hat{\imath}\, 2E_0 \cos(kz - \omega t)$$

whose amplitude and direction are time invariant thus linearly polarized. The most generic mathematical representation for the state of light polarization is given when $E_{0x} \neq E_{0y}$ and $\theta_{phase}$ being some arbitrary angle. Thus:

$$E_x = E_{0x}\cos(kz - \omega t) \rightarrow \frac{E_x}{E_{0x}} = \cos(kz - \omega t) \rightarrow \sin(kz - \omega t) = \left[1 - \left(\frac{E_x}{E_{0x}}\right)^2\right]^{1/2}$$

$$E_y = E_{0y}\cos(kz - \omega t + \theta_{phase}) \rightarrow \frac{E_y}{E_{0y}} = \cos(kz - \omega t)\cos(\theta_{phase}) - \sin(kz - \omega t)\sin(\theta_{phase})$$

$$\rightarrow \left(\frac{E_y}{E_{0y}} - \frac{E_x}{E_{0x}}\cos(\theta_{phase})\right)^2 = \left[1 - \left(\frac{E_x}{E_{0x}}\right)^2\right]\sin^2(\theta_{phase})$$

$$\rightarrow \left(\frac{E_y}{E_{0y}}\right)^2 + \left(\frac{E_x}{E_{0x}}\right)^2 - 2\left(\frac{E_y}{E_{0y}}\right)\left(\frac{E_x}{E_{0x}}\right)\cos(\theta_{phase}) = \sin^2(\theta_{phase}) \quad (2.1\text{-}1)$$

Which is the most generic equation of an ellipse that makes an angle $\varphi$ with $E_x$ axis where:

$$\tan 2\varphi = \frac{2E_{0x}E_{0y}\cos(\theta_{phase})}{E_{0x}^2 - E_{0y}^2}$$

In case of $\varphi = 0$ or $\theta_{\text{phase}} = \pm\pi/2$, the resulting expression for elliptical polarization, i.e. $E$-state, takes a more familiar form:

$$\left(\frac{E_y}{E_{0y}}\right)^2 + \left(\frac{E_x}{E_{0x}}\right)^2 = 1 \quad (2.1\text{-}2)$$

It is obvious when $E_{0x} = E_{0y}$, an equation of circle $E_x^2 + E_y^2 = E_0^2$ would result from Equation (2.1-2) representing $R$- or $L$-states. It can also be shown that for even and odd multiples of $\theta_{phase} = \pi$, Equation (2.1-1) leads to $E_y = \frac{E_{0y}}{E_{0x}} E_x$ and $E_y = -\frac{E_{0y}}{E_{0x}} E_x$ both of which are linear equation, hence $P$-states of polarization. This is important in understanding the principle governing the changes in polarization, i.e. from $P$-state to $E$-state, as light propagates in a magnetic media with anisotropic dielectric constant. Ellipticity is defined as the semiminor/semimajor axes ratio, i.e. $b/a$ as seen in Figure 1. This is an undesirable effect (with respect to this report), cause by circular dichroism which is responsible for distortion of the original EM wave. When light with $P$-state of polarization, (composed of $R$- and $L$-states with equal amplitudes), passes through a dichroic media with complex indices of refraction $N_\pm = n_\pm - ik_\pm$,

LCP and RCP lights experience different absorptions due to two different extinction coefficients $k_\pm$, resulting in different amplitudes leading to $E$-state of polarization.

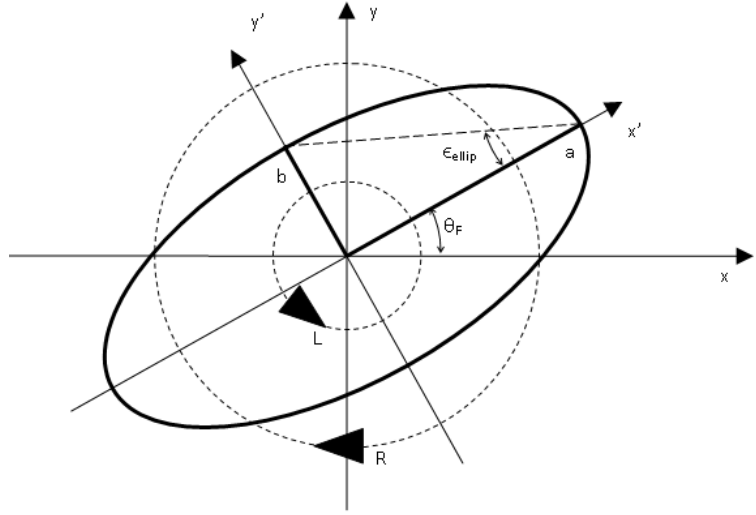

**Figure 1: Depiction of dichroic and birefringence distortions on a linearly polarized light.**

## 2.2　Faraday rotation

Faraday Rotation (FR) is the result of birefringence, i.e. two different indices of refraction $n_\pm$, experienced by RCP and LCP lights. This leads to two different optical paths. As a result, one will lag the other causing rotation of polarization plane, see Figure 1.

A brief mathematical treatment governing FR in thin films shall be presented here. According to classical optics, the total resultant amplitude transmission coefficient due to multiple reflections for a dielectric layer $L$, on a substrate $S$ when light is incident at right angle from the air $A$ is given by:

$$t_{ALS} = \frac{t_{AL}t_{LS}e^{-i\gamma Nd}}{1 - r_{LA}r_{LS}e^{-2i\gamma Nd}} = \frac{t_{AL}t_{LS}e^{-i\gamma Nd}}{1 + r_{AL}r_{LS}e^{-2i\gamma Nd}}$$

Where $N = n - ik$ is the complex refractive index, $t_{AL}$ and $t_{LS}$ are the transmission coefficients, $r_{LA}$ and $r_{LS}$ are the reflection coefficients at air-layer and layer-substrate interfaces, $d$ is the thickness of the layer and $\gamma = 2\pi/\lambda$ where $\lambda$ is the wavelength of incident light. Furthermore,

$$r_{AL} = \frac{1-N}{1+N} \quad AND \quad r_{LS} = \frac{N-s}{N+s}$$

$$t_{AL} = 1 + r_{AL} = \frac{2}{1+N} \quad AND \quad t_{LS} = 1 + r_{LS} = \frac{2N}{N+s}$$

Here, *s* is the refractive index of the substrate. The most generic mathematical representation for Faraday's rotation for a single magneto-optic layer on a substrate is given by Višňovský[8] as:

$$\theta_F = i \frac{\partial t_{ALS}}{\partial N} \frac{\Delta N}{t_{ALS}}$$

where $\Delta N = \frac{1}{2}(N_+ - N_-)$ is the complex circular birefringence and $N_\pm = n_\pm - ik_\pm$ are the complex indices of refraction experienced by RCP and LCP light as they propagate through the magnetic layer. Therefore, an analytical expression in terms of all physical quantities that playing a role in Faraday rotation may be derived as follow:

$$\rightarrow t_{ALS} = \frac{\frac{2}{1+N}\frac{2N}{N+s}e^{-i\gamma N d}}{1-\left(\frac{N-1}{N+1}\right)\left(\frac{N-s}{N+s}\right)e^{-2i\gamma N d}} = \frac{\frac{4N}{N^2+Ns+N+s}e^{-i\gamma N d}}{\frac{N^2+Ns+N+s}{N^2+Ns+N+s}-\frac{N^2-Ns-N+s}{N^2+Ns+N+s}e^{-2i\gamma N d}}$$

$$= \frac{4Ne^{-i\gamma N d}}{N^2+Ns+N+s-(N^2-Ns-N+s)e^{-2i\gamma N d}}$$

$$\text{let } f = 4Ne^{-i\gamma N d}$$

$$\rightarrow f' = 4e^{-i\gamma N d}(1-iN\gamma d)$$

$$\text{let } g = N^2+Ns+N+s-(N^2-Ns-N+s)e^{-2i\gamma N d}$$

$$\rightarrow g' = 2N+s+1-\left[(2N-s-1)e^{-2i\gamma N d}-2i\gamma d(N^2-Ns-N+s)e^{-2i\gamma N d}\right]$$

$$\rightarrow \frac{\partial t_{ALS}}{\partial N} = \frac{f'g-g'f}{g^2}$$

$$= \frac{(1-iN\gamma d)[N^2+Ns+N+s-(N^2-Ns-N+s)e^{-2i\gamma N d}]-\left[2N+s+1-[(2N-s-1)e^{-2i\gamma N d}-2i\gamma d(N^2-Ns-N+s)e^{-2i\gamma N d}]\right]N}{[N^2+Ns+N+s-(N^2-Ns-N+s)e^{-2i\gamma N d}]} \times \frac{1}{N}$$

$$= \frac{(1-iN\gamma d)\left[\frac{N^2+Ns+N+s}{N^2+Ns+N+s}-\frac{(N^2-Ns-N+s)}{N^2+Ns+N+s}e^{-2i\gamma N d}\right]-\frac{2N^2+Ns+N}{N^2+Ns+N+s}+\frac{(2N^2-Ns-N)}{N^2+Ns+N+s}e^{-2i\gamma N d}-\frac{2i\gamma d(N^2-Ns-N+s)}{N^2+Ns+N+s}Ne^{-2i\gamma N d}}{\left[\frac{N^2+Ns+N+s}{N^2+Ns+N+s}-\frac{(N^2-Ns-N+s)}{N^2+Ns+N+s}e^{-2i\gamma N d}\right]} \times \frac{1}{N}$$

$$\text{note: } \frac{N^2-Ns-N+s}{N^2+Ns+N+s} = -r_{AL}r_{LS} \quad \text{AND} \quad \frac{r_{LA}+r_{LS}}{2} = \frac{N^2-s}{N^2+Ns+N+s}$$

$$\rightarrow \frac{\partial t_{ALS}}{\partial N}\frac{1}{t_{ALS}} = \frac{1+r_{AL}r_{LS}e^{-2i\gamma N d}-iN\gamma d-iN\gamma d r_{AL}r_{LS}e^{-2i\gamma N d}-\left(1+\frac{r_{LA}+r_{LS}}{2}\right)+\left(-r_{AL}r_{LS}+\frac{r_{LA}+r_{LS}}{2}\right)e^{-2i\gamma N d}+2iN\gamma d r_{AL}r_{LS}e^{-2i\gamma N d}}{1+r_{AL}r_{LS}e^{-2i\gamma N d}} \times \frac{1}{N}$$

$$= \frac{(r_{LA} + r_{LS})(e^{-2i\gamma Nd} - 1) + i2N\gamma d(r_{AL}r_{LS}e^{-2i\gamma Nd} - 1)}{1 + r_{AL}r_{LS}e^{-2i\gamma Nd}} \times \frac{1}{2N}$$

$$\text{note: } r_{AL} = -r_{LA}$$

$$\rightarrow \theta_F = \frac{2N\gamma d(r_{LA}r_{LS}e^{-2i\gamma Nd} + 1) + i(e^{-2i\gamma Nd} - 1)(r_{LA} + r_{LS})}{1 - r_{LA}r_{LS}e^{-2i\gamma Nd}} \times \frac{\Delta N}{2N} \qquad (2.2\text{-}1)$$

## 2.3  Origin of Magneto-optic Effect

In general, MO effects are due to electric dipole transitions in atoms constituting the material at the presence of a magnetic field. When a linearly polarized light impinges on a magneto-optic material in z direction, electric vector of the two constituent lights, i.e. RCP and LCP, set elastically bound electrons into circular motion in *x-y* plane which leads to transition from ground state $\langle g|$ to excited state $|e\rangle$. Since RCP and LCP light have two opposite spin angular momentum, $\pm\hbar$, electrons driven by RCP end up with opposite angular momentum to those driven by LCP light, however, they maintain equal electric dipole moments since their displacement from the center of circular path is equal. In the presence of magnetic field in the *z*-direction, i.e. either external or internal due to magnetization, LCP driven electrons that are set in circular motion counter clockwise, experience an inward radial force towards the center of the circle. RCP driven electrons, set in circular motion clockwise, experience an outward radial force. Consequently, in the presence of magnetic field there will be two distinct values for displacement *x*, dipole moment $p = xq$, electric polarization $\mathbf{P} = Np$, permittivity $\varepsilon = 1 + \mathbf{P}/\varepsilon_0\mathbf{E}$ and hence excited states for LCP and RCP lights. Change in orbital momentum by magnetic field, such as the one described above, is the main factor in diamagnetism[9], see Figure 2. This is rather a simplistic recount of split in electronic state. Exact mechanism is not fully understood as spin-orbit coupling (i.e. paramagnetic effect), *e-e* interactions, exchange interactions, phonon interaction and other phenomena all play a role in the magneto-optic effect.

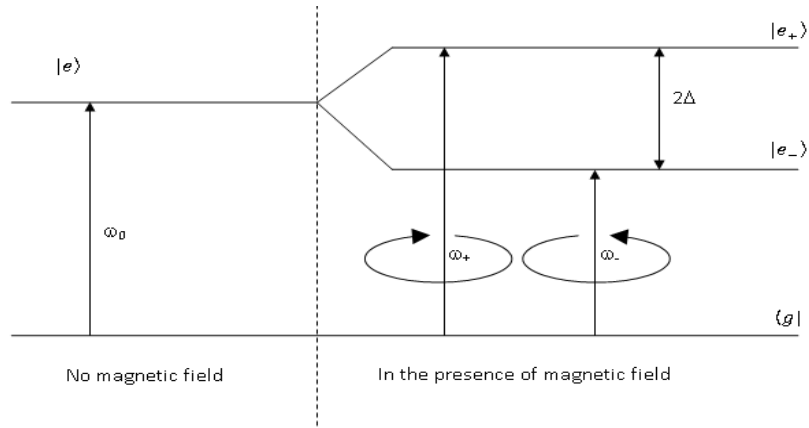

**Figure 2: Diamagnetic effect**

## 2.4 Microscopic model

Electric displacement induced by the electric field component of the EM wave is described by generic expressions:

$$D_x = \varepsilon_{xx} E_x + \varepsilon_{xy} E_y + \varepsilon_{xz} E_z$$

$$D_y = \varepsilon_{yx} E_x + \varepsilon_{yy} E_y + \varepsilon_{yz} E_z$$

$$D_z = \varepsilon_{zx} E_x + \varepsilon_{zy} E_y + \varepsilon_{zz} E_z$$

Or

$$\begin{bmatrix} D_x \\ D_y \\ D_z \end{bmatrix} = \begin{bmatrix} \varepsilon_{xx} & \varepsilon_{xy} & \varepsilon_{xz} \\ \varepsilon_{yx} & \varepsilon_{yy} & \varepsilon_{yz} \\ \varepsilon_{zx} & \varepsilon_{zy} & \varepsilon_{zz} \end{bmatrix} \begin{bmatrix} E_x \\ E_y \\ E_z \end{bmatrix}$$

For magneto-optic material magnetized in the *z*-direction where light is impinging on the surface at right angle, i.e. in *z*-direction, $\varepsilon_{yz} = \varepsilon_{xz} = \varepsilon_{zx} = \varepsilon_{zy} = 0, \varepsilon_{xx} = \varepsilon_{yy}$ and $\varepsilon_{yx} = -\varepsilon_{xy}$. The resultant dielectric tensor is then:

$$\hat{\varepsilon} = \begin{bmatrix} \varepsilon_{xx} & \varepsilon_{xy} & 0 \\ -\varepsilon_{xy} & \varepsilon_{xx} & 0 \\ 0 & 0 & \varepsilon_{zz} \end{bmatrix} \qquad (2.4\text{-}1)$$

Macroscopic optical properties such as complex refractive index $N = n - ik = \sqrt{\varepsilon_{xx}}$, thus refractive index $n(\lambda) = \Re\sqrt{\varepsilon_{xx}}$, extinction coefficient $k(\lambda) = -\Im\sqrt{\varepsilon_{xx}}$ and circular birefringence indices of refraction $N_\pm = n_\pm - ik_\pm = \sqrt{\varepsilon_{xx} \pm i\varepsilon_{xy}}$ are all functions of diagonal and off-diagonal dielectric constants. Expanding the definition of complex circular birefringence, $\Delta N = \frac{1}{2}(N_+ - N_-)$ leads to:

$$\Delta N = \frac{1}{2}\left(\sqrt{\varepsilon_{xx} + i\varepsilon_{xy}} - \sqrt{\varepsilon_{xx} - i\varepsilon_{xy}}\right) = \frac{1}{2}\left(\frac{i2\varepsilon_{xy}}{\sqrt{\varepsilon_{xx} + i\varepsilon_{xy}} + \sqrt{\varepsilon_{xx} - i\varepsilon_{xy}}}\right)$$

For $\{\varepsilon_{xy} \ll \varepsilon_{xx}\}$

$$\Delta N \approx \frac{i\varepsilon_{xy}}{2\sqrt{\varepsilon_{xx}}} = \frac{i\varepsilon_{xy}}{2N}$$

Substituting Δ*N* in Equation (2.2-1) results in:

$$\theta_F = \Re\left\{\frac{\left(\frac{\pi d}{\lambda}\right)\left(1 + r_{LA}r_{LS}e^{-\frac{i4\pi Nd}{\lambda}}\right) + \frac{i\left(e^{-\frac{i4\pi Nd}{\lambda}} - 1\right)(r_{LA} + r_{LS})}{4N}}{1 - r_{LA}r_{LS}e^{-\frac{i4\pi Nd}{\lambda}}} \times \frac{i\varepsilon_{xy}}{N}\right\}$$

Furthermore, substituting $N$ by $\sqrt{\varepsilon_{xx}}$ leads to:

$$\theta_F(\lambda) = \Re\left\{\frac{i\varepsilon_{xy}}{\sqrt{\varepsilon_{xx}}} \left(\frac{\left(\frac{\pi d}{\lambda}\right)\left(1 + r_{fa}r_{fs}e^{-\frac{i4\pi d\sqrt{\varepsilon_{xx}}}{\lambda}}\right) + i\frac{\left(e^{-\frac{i4\pi d\sqrt{\varepsilon_{xx}}}{\lambda}} - 1\right)(r_{fa} + r_{fs})}{4\sqrt{\varepsilon_{xx}}}}{\left(1 - r_{fa}r_{fs}e^{-\frac{i4\pi d\sqrt{\varepsilon_{xx}}}{\lambda}}\right)}\right)\right\} \quad (2.4\text{-}2)$$

This makes FR a function of displacement due to off-diagonal element of dielectric tensor. It is therefore, imperative to have as exact mathematical expression as possible for the two quantities $\varepsilon_{xx}$ and $\varepsilon_{xy}$.

Quantum mechanical treatment of electric dipole interaction with light for the scenario described above has led to the following expressions for diagonal and off-diagonal elements of dielectric tensor [10]:

$$\varepsilon_{xx} = 1 + \omega_P^2 \sum_{transitions} \sum_{\pm} f_{\pm} \frac{\omega_{0\pm}^2 - \omega^2 + \Gamma^2 - i2\omega\Gamma}{\left(\omega_{0\pm}^2 - \omega^2 + \Gamma^2\right)^2 + (2\omega\Gamma)^2} \quad (2.4\text{-}3)$$

$$\varepsilon_{xy} = i\frac{\omega_P^2}{2} \sum_{transitions} \sum_{\pm} (\pm 1) \frac{f_{\pm}}{\omega_{0\pm}} \frac{\omega(\omega_{0\pm}^2 - \omega^2 - \Gamma^2) - i\Gamma(\omega_{0\pm}^2 + \omega^2 + \Gamma^2)}{\left(\omega_{0\pm}^2 - \omega^2 + \Gamma^2\right)^2 + (2\omega\Gamma)^2} \quad (2.4\text{-}4)$$

Here $\omega_p$, $\omega_0$ and $\omega$ are the plasma, resonant transition and incident light frequencies, $\Gamma$ is the half line-width of the transition between the ground state $\langle g|$ and the two excited states $|e_{\pm}\rangle$, $2\Delta$ is the split between the two excited states, $\omega_{0\pm} = \omega_0 \pm \Delta$ and $f_{\pm} \approx \frac{f}{2}\left(1 \pm \frac{\Delta}{\omega_0}\right)$, are the oscillator's strength associated to the left- and right-handed circular polarizations resulting from the two excited states. Assuming single resonance transition at $\omega = \omega_0$, $\omega_0 \ll \Delta$ and $\omega_0 \gg \Gamma$, Equations (2.4-3) and (2.4-4) can be approximated to:

$$n^2 \approx 1 + \frac{\Omega^2}{1 - \Lambda^2} \quad AND \quad (2.4\text{-}5)$$

$$k \approx \Lambda \frac{\Gamma}{\omega_0}\left(\frac{\Omega}{1 - \Lambda^2}\right)^2 \sqrt{\left(\frac{1 - \Lambda^2}{1 - \Lambda^2 + \Omega^2}\right)} \quad where \quad \Omega = \frac{\omega_p\sqrt{f}}{\omega_0} \quad and \quad \Lambda = \frac{\lambda_0}{\lambda} \quad (2.4\text{-}6)$$

Note that equation (2.4-5) is a Sellmeier equation for dispersion.

## 2.4.1 Obtaining *n, k*, $\varepsilon_{xx}$ and $\varepsilon_{xy}$

To obtain dispersive values for dielectric constants, $\varepsilon_{xx}$, one needs to know the refractive index *n* and the extinction coefficient *k*, for a given material. A technique proposed by Swanepoel [11], allows extraction of *n* and *k* from interference fringes in the transmission spectrum of a thin film similar to that shown in Figure 3. The spectrum is divided in 4 regions: ***transparent***, ***weak***, ***medium*** and ***strong*** absorptions, each of which has its own remedy for calculating *n* and *k*. The idea is to calculate *n* and *k* at different wavelengths which then can be fitted into a dispersion relation like that of Sellmeier's formula. Prior to calculations, maxima and minima fringes of the transmission spectrum must be connect with smooth curves $T_M$ and $T_m$. Values at each wavelength must be read on these curves rather than the actual spectrum. Every maximum, $t_M$, has its own corresponding minima, $t_m$, at a particular wavelength, see marked positions on $T_M$ and $T_m$ curves.

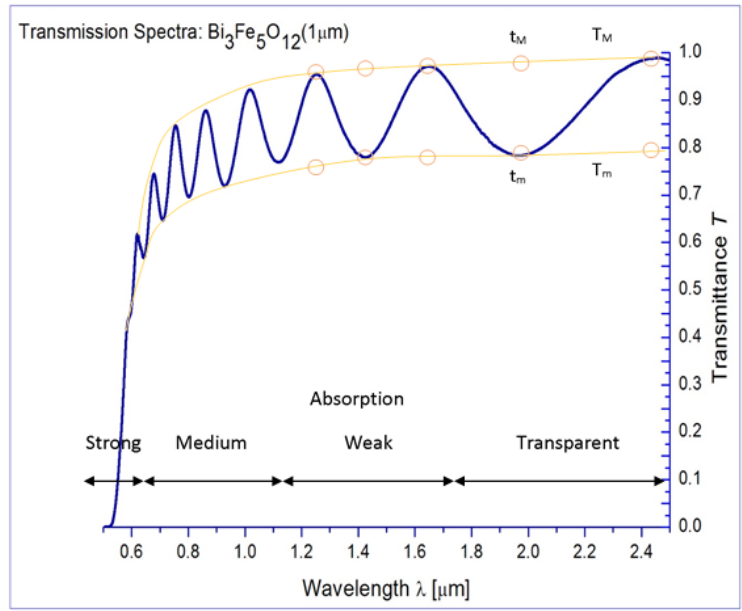

**Figure 3: Transmission spectra of a BIG 1µm film**

In ***transparent*** region refractive index *n* and extinction coefficients *k* are given by:

$$n = \sqrt{\left[M + \sqrt{(M^2 - s^2)}\right]} \; where \; M = \frac{2s}{t_m} - \frac{s^2 + 1}{2} \; and \; k = 0$$

Here, *s* is the refractive index of the substrate.

In regions with ***weak*** and ***medium*** absorption:

$$n = \sqrt{\left[M + \sqrt{(M^2 - s^2)}\right]} \quad where \quad M = 2s\frac{t_M - t_m}{t_m t_M} + \frac{s^2 + 1}{2}$$

$$x = \frac{E_M - \sqrt{[E_M^2 - (n^2 - 1)^3(n^2 - s^4)]}}{(n - 1)^3(n - s^2)} \quad where \quad E_M = \frac{8n^2 s}{t_M} + (n^2 - 1)(n^2 - s^2)$$

Once $x$ is known, following equations may be used to calculate $k$

$$x = \exp(-\alpha d) \quad where \quad \alpha = 4\pi k/\lambda$$

In region with **strong** absorption, $n$ and $k$ may not be calculated independently, consequently, Swanepoel suggests extrapolating $n$ from values previously calculated, and then use the expression for $x$ to calculate $k$.

Alternatively, if the thickness of the thin film is reliably known, one can simply apply $m\lambda=2nd$ at extrema positions to extract $n$ ignoring $k$. Here $d$ is the thickness of the thin film and $m$ is an integer for maxima and half integer for minima identifying a particular interference fringe.

Once $n$ and $k$ values are extracted, they can be fitted into Sellmeier's Equation (2.4-5) and (2.4-6) to identify the number of dipole transitions and their associated material parameters, $\omega_0, \omega_p\sqrt{f}, \Gamma$ and $\Delta$. Once these parameters are known, $n$ and $k$ must be reconstructed from $\varepsilon_{xx}$ and $\varepsilon_{xy}$ using Equations (2.4-3) and (2.4-4) in order to simulate the transmission spectrum using the following Swanepoel formula:

$$T = \frac{Ax}{B - Cx + Dx^2} \tag{2.4-7}$$

$$A = 16s(n^2 + k^2)$$

$$B = [(n + 1)^2 + k^2][(n + 1)(n + s^2) + k^2]$$

$$C = [(n^2 - 1 + k^2)(n^2 - s^2 + k^2) - 2k^2(s^2 + 1)]2\cos\varphi \\ - k[2(n^2 - s^2 + k^2) + (s^2 + 1)(n^2 - 1 + k^2)]2\sin\varphi \tag{2.4-8}$$

$$D = [(n - 1)^2 + k^2][(n - 1)(n - s^2) + k^2]$$

$$where \quad \varphi = \frac{4\pi nd}{\lambda} \quad and \quad x = \exp\left(-\frac{4\pi kd}{\lambda}\right),$$

Simulated spectrum must be a close match to the experimentally obtained spectrum. Otherwise further fine adjustments to the material parameters are needed. This process must be repeated until simulated spectrum closely matches experiment. FR should also be simulated using Equation (2.4-2) and compared to the experimental data as a double check.

## 2.4.2 4×4 Matrix Formalism and MOPC

Previous section dealt with the process of obtaining diagonal and off-diagonal dielectric tensor elements, $\varepsilon_{xx}$ and $\varepsilon_{xy}$. In this section I describe how one can model a one dimensional MOPC whose layers are composed of MO material for which those quantities are known. Transmission spectrum, FR and Ellipticity of a one dimensional layered structures such as MOPC, can be determined from the elements of the resultant transmission matrix. Višňovský et al [12] developed a 4x4 matrix formalism for a layered structure composed of N layers, including the semi-infinite substrate, separated by N+1 interfaces. Relation between the electric field amplitudes at interface *j-1* and *j* is governed by:

$$\vec{E}^{j-1} = T_{j-1,j}\vec{E}^j \rightarrow \begin{bmatrix} E_1^{j-1} \\ E_2^{j-1} \\ E_3^{j-1} \\ E_4^{j-1} \end{bmatrix} = \begin{bmatrix} T_{11}^{j-1,j} & T_{12}^{j-1,j} & 0 & 0 \\ T_{21}^{j-1,j} & T_{22}^{j-1,j} & 0 & 0 \\ 0 & 0 & T_{33}^{j-1,j} & T_{34}^{j-1,j} \\ 0 & 0 & T_{43}^{j-1,j} & T_{44}^{j-1,j} \end{bmatrix} \begin{bmatrix} E_1^j \\ E_2^j \\ E_3^j \\ E_4^j \end{bmatrix}$$

Where $T_{j-1,j}$ is the transmission matrix relating the electric field amplitudes at interfaces *j-1* and *j*. With layers being magnetized parallel to *z* axis and light being incident at right angle to the surface in positive *z* direction, $E_1^j$ and $E_3^j$ denote the complex amplitudes of two circularly polarized (CP) lights propagating in the positive z direction whereas $E_2^j$ and $E_4^j$ denote the complex amplitudes of two CP lights propagating in the negative z direction. Then the elements of the T-matrix are given by:

$$T_{11}^{j-1,j} = \frac{1}{2N_+^{j-1}}\left(N_+^{j-1} + N_+^j\right)e^{i\beta_+^j}$$

$$T_{12}^{j-1,j} = \frac{1}{2N_+^{j-1}}\left(N_+^{j-1} - N_+^j\right)e^{-i\beta_+^j}$$

$$T_{21}^{j-1,j} = \frac{1}{2N_+^{j-1}}\left(N_+^{j-1} - N_+^j\right)e^{i\beta_+^j}$$

$$T_{22}^{j-1,j} = \frac{1}{2N_+^{j-1}}\left(N_+^{j-1} + N_+^j\right)e^{-i\beta_+^j}$$

$$T_{33}^{j-1,j} = \frac{1}{2N_-^{j-1}}\left(N_-^{j-1} + N_-^j\right)e^{i\beta_-^j}$$

$$T_{34}^{j-1,j} = \frac{1}{2N_-^{j-1}}\left(N_-^{j-1} - N_-^j\right)e^{-i\beta_-^j}$$

$$T_{43}^{j-1,j} = \frac{1}{2N_-^{j-1}}\left(N_-^{j-1} - N_-^j\right)e^{i\beta_-^j}$$

$$T_{44}^{j-1,j} = \frac{1}{2N_-^{j-1}}(N_-^{j-1} + N_-^j)e^{-i\beta_-^j}$$

Where $N_\pm^j = \sqrt{\varepsilon_{xx}^j \pm i\varepsilon_{xy}^j}$, $\beta_\pm^j = \frac{2\pi}{\lambda}N_\pm^j d^j$ and $d^j$ is the thickness of the $j^{th}$ layer. The total resultant transmission matrix for a MOPC is then given by $\boldsymbol{M} = \prod_{j=1}^{N+1} \boldsymbol{T}_{j-1,j}$ thus $\vec{E}^0 = \boldsymbol{M}\vec{E}^{N+1}$. Note that as light enters the MOPC from one side and travels through each layer, $E_2^j$ and $E_4^j$ may be non-vanishing at each interface except at the exit interface for which there will be no reflection. This implies that, at substrate-air interface $E_2^{N+1} = E_4^{N+1} = 0$ which leads to:

$$\begin{bmatrix} E_+^i \\ E_+^r \\ E_-^i \\ E_-^r \end{bmatrix} = \begin{bmatrix} M_{11} & M_{12} & 0 & 0 \\ M_{21} & M_{22} & 0 & 0 \\ 0 & 0 & M_{33} & M_{34} \\ 0 & 0 & M_{43} & M_{44} \end{bmatrix} \begin{bmatrix} E_+^t \\ 0 \\ E_-^t \\ 0 \end{bmatrix} \quad (2.4\text{-}9)$$

Here $i$ denotes 'incident', $t$ 'transmitted' and $r$ 'reflected'. However, since we only deal with transmission matrix elements in our simulation, it is not possible to set $E_2^{N+1} = E_4^{N+1} = 0$. Instead we shall set $T_{12}=T_{22}=T_{34}=T_{44}=0$ when constructing the $T$-matrix for the substrate-air interface only. Consequence of failing to do so results in closely packed unwanted oscillation all along the spectra curves. Once $\boldsymbol{M}$ matrix is obtained, magneto-optic effects of the MOPC can be calculated from its elements as follow:

$$Transmittance \quad T(\lambda) = \frac{1}{2}(|M_{11}|^{-2} + |M_{33}|^{-2}) \quad (2.4\text{-}10)$$

$$Faraday's\ rotation \quad \Theta_F(\lambda) = -\frac{1}{2}\arg(\chi) \quad (2.4\text{-}11)$$

$$ellipticity \quad \tan\epsilon_{ellip}(\lambda) = \frac{|\chi| - 1}{|\chi| + 1} \quad (2.4\text{-}12)$$

Where: $\chi(\lambda) = \frac{M_{11}}{M_{33}}$ and $\chi(\lambda) = -i\frac{M_{11}-M_{33}}{M_{11}+M_{33}}$ when incident light is linearly polarized at 45° (i.e. $\frac{E_-^i}{E_+^i} = 1$) and at 0° (i.e. $\frac{E_-^i}{E_+^i} = 0$) respectively[13].

# 3. Magneto-optic Material

## 3.1 Garnets

In general, garnets are identified by their unit formula

$$\{C_3^{3+}\}[A_2^{3+}](D^{3+}O_4^{2-})_3$$

Here {C}, [A] and (D) represent trivalent cations occupying dodecahedron (c) sites, octahedron (a) sites and tetrahedron (d) sites within the garnet's unit cell and '$O$' stands for Oxygen, see Figure 4. There are 8 formula units within a unit cell. Garnet configuration may be generalized as follow[14] [15] :

1. There are 24 (d) sites in a unit cell where each (d) site contains a $D^{3+}$ ions tetrahedrally surrounded by 4 $O^{2-}$ ions.
2. There 16 (a) sites in a unit cell each containing a $A^{3+}$ ions octahedrally surrounded by 6 $O^{2-}$ ions.
3. There are 24 (c) sites per unit cell each containing a $C^{3+}$ ion surrounded by 8 $O^{2-}$ ions in a dodecahedral (or icosahedral) configuration.

Flexibility of garnets in accepting large number of various ions at each site, allows one to tailor its physical properties for specific purpose. One such variation is called rare earth garnets with a general formula unit of $\{Re_3^{3+}\}[A_2^{3+}](A^{3+}O_4^{2-})_3$ where 'Re' stands for any of the rare earth elements occupying (c) sites, and 'A' stands for $Al^{3+}, Ga^{3+}$ and $Fe^{3+}$ occupying (a) and (d) sites. Magnetic rare earth garnets are the result of $Fe^{3+}$ presence.

Yttrium iron garnet (YIG), a type of magnetic rare earth garnet, is a synthetic ferrimagnetic crystal identified by its unit formula $Y_3Fe_2(FeO_4)_3$ having a cubic lattice structure with near perfect cubic symmetry having lattice constant a = 12.376 Å [14]. Since early days of its discovery, YIG has been used in magnetic and crystallographic studies. Its crystallographic structure has been generalized as a prototype for synthesizing large number of other magnetic garnets with different compositions. Origin of ferrimagnetism, (or anti-ferromagnetism), in such garnets are due to the superexchange interaction between the $Fe^{3+}$ ions occupying the (a) and (d) sites via $O^{2-}$ ions producing a total magnetic moment per formula unit of 5 $\mu_B$ [14], where $\mu_B = \frac{e\hbar}{2m_e}$ is the Bohr magneton.

Completely substituted bismuth iron garnet $Bi_3^{3+}Fe_2^{3+}(Fe^{3+}O_4^{2-})_3$, (BIG), has the highest specific FR to date whereas rare earth garnet $Gd_3^{3+}Ga_2^{3+}(Ga_2^{3+}O_4^{2-})_3$, (GGG), is highly transparent, i.e. minimum insertion loss. Therefore we limit our choices to BIG/GGG multilayer structures as a model for now. Lattice constant for BIG is found to be 12.627 Å [2]. Therefore, BIG has a density, in terms of formula units/cm³, of $\frac{8}{(12.627\times 10^{-8})^3} = 3.973 \times 10^{21}$ formula units/cm³. Similarly, GGG with lattice constant of 12.383 Å [15] has a density of $4.213 \times 10^{21}$ formula units/cm³. These quantities will be used in later chapters.

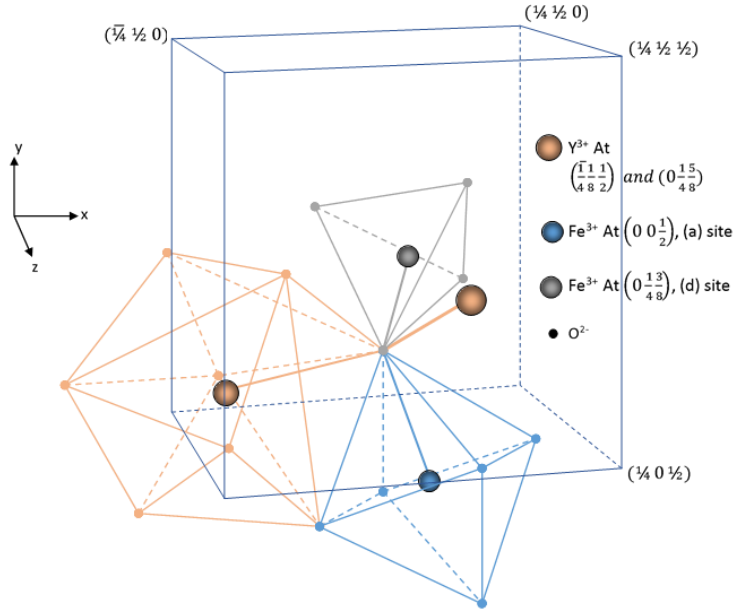

**Figure 4: YIG lattice. Inspired from [16]**

# 4. Amplifying Magneto-Optical Photonic Crystal

## 4.1   Innovative Step

Ultimate goal of the work undertaken in this thesis is the design of a multilayer structure and determination of its constituents material, such as erbium doped rare earth magnetic/non-magnetic garnets, e.g. Er:BIG/Er:GGG, capable of producing 45° FR, transmittance higher than 0.7 and minimal ellipticity at optimum total film thickness. Thicknesses of garnet layers are chosen to be an odd integer of a quarter resonance wavelength $\lambda_s/4n(\lambda_s)$ in Bragg reflectors and integer of a half resonance wavelength $\lambda_s/2n(\lambda_s)$ in microcavities. Here $n(\lambda_s)$ is the refractive index of corresponding garnet materials at the resonance wavelength $\lambda_s$. Amplifying Magneto-Optical Photonic Crystals, (AMOPC), can be used as high performance optical amplifiers, optical isolators, magnetic field and current controlled modulators. Being used as optical isolator, see Figure 5, AMOPCs, 3, has a signal light laser 1, a polarizer 2, a pumping light laser 4, a thin film made of epitaxially grown multilayered Er-doped all-garnet film 3.1-3.3, a substrate made of a garnet single crystal 3.4, an analyzer 5, and a photodetector 6. The film and the substrate both may form the amplifying element.

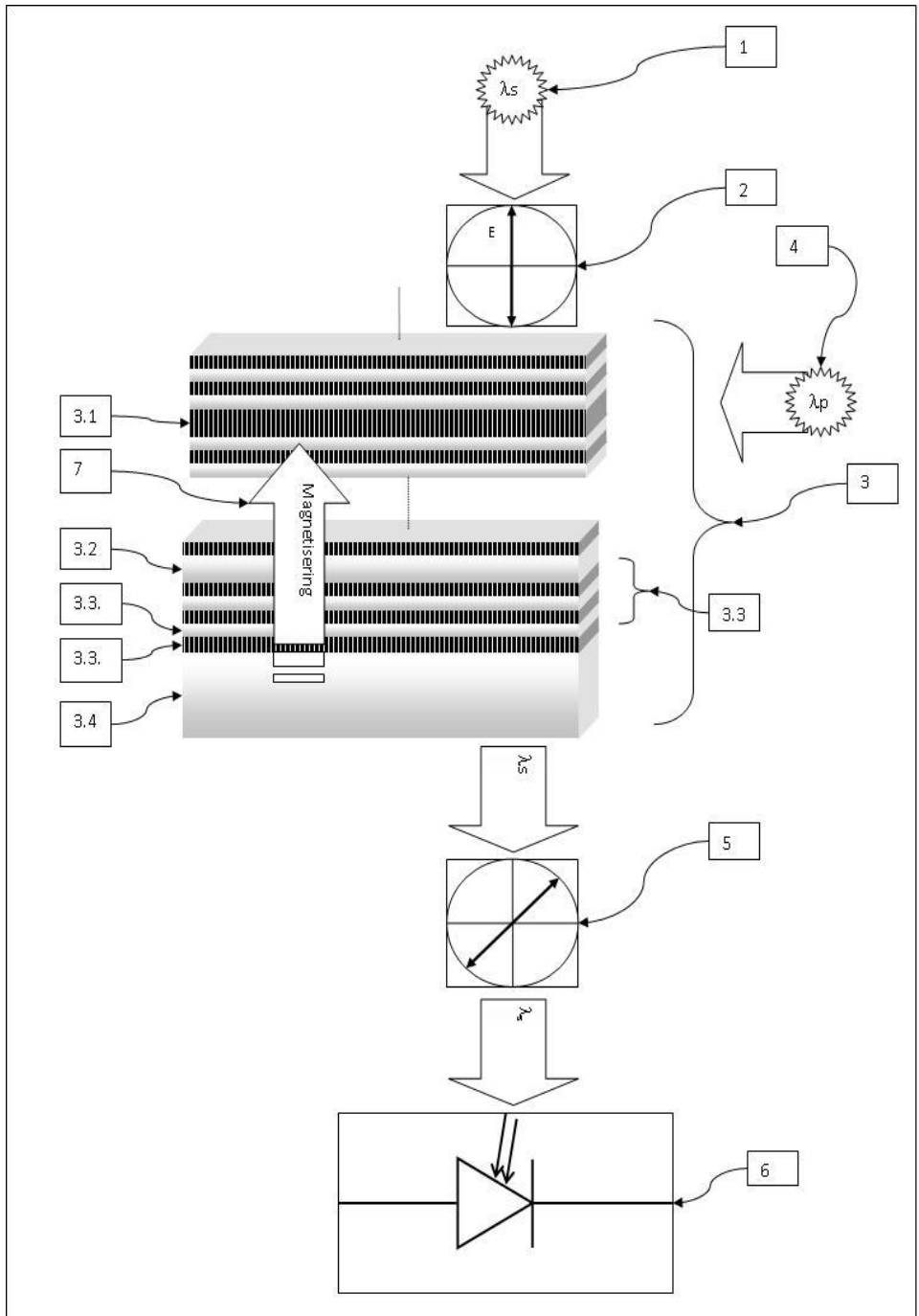

**Figure 5: A possible use of AMOPC as optical isolator**

## 4.2 Technical Background

Optically gain media necessitates Intensive luminescence condition. Recently, high room temperature photoluminescence (PL) was observed in several rare earth gallium and iron contained garnets [5] . When considering rare earth iron garnets, the $^4I_{11/2}$ energy level of $Er^{3+}$ and the $^4T_{1g}$ level of octahedrally coordinated ferric ions are nearly resonant in energy. $Fe^{3+}$ can be excited by a solid state 980 nm lasers whose light is absorbed at the narrow discrete band around $^4T_{1g}$ level under three level laser scheme. There are sixteen octahedrally coordinated Fe ions per one Er atom in the erbium substituted iron garnet unit cell. In PL process, the net $Fe^{3+}$ absorption cross section at 980 nm is 16 times higher than that of $Er^{3+}$. For C-band PL, Fe-promoted sensitizing effect occurs at Fe-concentrations above 2.5 formula units. Furthermore, Er-dopant did not change the giant FR present in $Bi_3Fe_5O_{12}$. $Bi_{2.9}Er_{0.1}Fe_5O_{12}$ film shows specific FR with a peak value of -30 deg/µm at 535 nm and equals -6.7 deg/µm and -1.63 deg/µm at 654 nm and 980 nm, respectively [5]. Therefore, doping the magnetic garnets poses no threat to the desired Faraday effect in our multilayer structure. Same is true for non-magnetic garnets.

Increase of Erbium content leads to luminescence increase unless effects of excitation quenching and precipitation of Erbium at high concentrations in garnet matrix impede further PL enhancement. Doubling Erbium-dopant has increased PL more than twice as it is observed in $Bi_{2.8}Er_{0.2}Fe_5O_{12}$ compared to $Bi_{2.9}Er_{0.1}Fe_5O_{12}$ [5]. Peshko et al, [17], showed that raising of Er concentration in $(Gd,Y)_3(Ga,Sc)_5O_{12}$ up to $x = 0.6$ formula units (3 atomic %) gradually increase of PL without any indication of quenching effect.

Under the 980 nm pumping radiation, a three-level scheme governs the laser transition in Er-doped garnets. Therefore, I consider a simplified three levels scheme where excited-state absorption to higher levels and upconversion processes are negligible (see Figure 6).

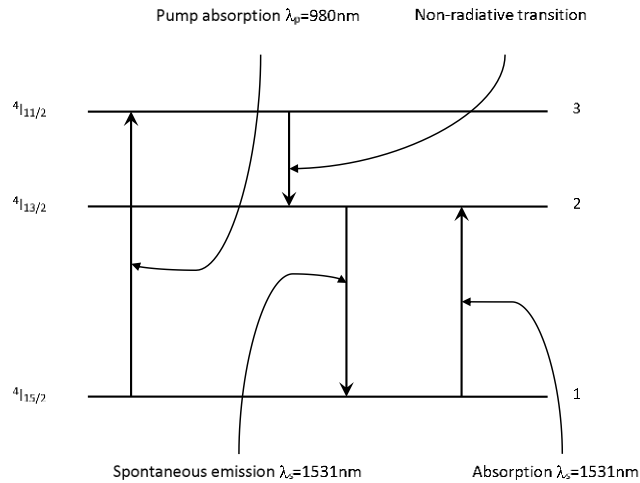

**Figure 6: Three level laser transition depicting $Er^{3+}$ energy states**

980 nm pumping radiation with the intensity $I_p$ excites electrons from the $^4I_{15/2}$ ground state 1 to the level 3 that is the upper-state of the $^4I_{11/2}$ Stark manifold. Non-radiative transition 3 → 2 occurs between the $^4I_{11/2}$ and $^4I_{13/2}$ levels. Spontaneous emission 2 → 1 competes with the absorption of a signal beam caused by the electron transition 1 → 2. Rate equations governing the dynamics of all the above mentioned processes are then:

$$\frac{\partial n_3}{\partial t} = \frac{\sigma_{13} I_p}{\hbar \omega_p} n_1 - \frac{\sigma_{31} I_p}{\hbar \omega_p} n_3 - \tau_{32}^{-1} n_3 - \tau_{31}^{-1} n_3$$

$$\frac{\partial n_2}{\partial t} = \frac{\sigma_{12} I_s}{\hbar \omega_s} n_1 - \frac{\sigma_{21} I_s}{\hbar \omega_s} n_2 - \tau_{21}^{-1} n_2 + \tau_{32}^{-1} n_3 \quad (4.2\text{-}1)$$

$$\frac{\partial n_1}{\partial t} = -\frac{\sigma_{13} I_p}{\hbar \omega_p} n_1 + \frac{\sigma_{31} I_p}{\hbar \omega_p} n_3 - \frac{\sigma_{12} I_s}{\hbar \omega_s} n_1 + \frac{\sigma_{21} I_s}{\hbar \omega_s} n_2 + \tau_{21}^{-1} n_2 + \tau_{31}^{-1} n_3$$

Here $n_i$ stands for the fractional level population of electron energy level $i$, $\tau_{ik}$ is the lifetime and $\sigma_{ik}$ is the cross-section, respectively, for the corresponding i → k transition. $I_p$ is the intensity of pumping beam and $I_s$ is the intensity of signal, $\hbar \omega_p$ and $\hbar \omega_s$ are the corresponding photons energy. Due to the multi-phonon non-radiative processes, $^4I_{11/2}$ lifetime is very short, i.e. $\tau_{32} \to 0$, therefore $n_3 \to 0$ and system of Equations (4.2-1) can be reduced to one equation that in the case of steady state becomes:

$$\frac{\partial n_1}{\partial t} = -\frac{\sigma_{13} I_p}{\hbar \omega_p} n_1 - \frac{\sigma_{12} I_s}{\hbar \omega_s} n_1 + \frac{\sigma_{21} I_s}{\hbar \omega_s} n_2 + \tau_{21}^{-1} n_2 = 0$$

$$\to -n_1 \left( \frac{\sigma_{13} I_p}{\hbar \omega_p} + \frac{\sigma_{12} I_s}{\hbar \omega_s} \right) + n_2 \left( \frac{\sigma_{21} I_s}{\hbar \omega_s} + \tau_{21}^{-1} \right) = 0$$

where $n_1 + n_2 = 1$

$$-\left( \frac{\sigma_{13} I_p}{\hbar \omega_p} + \frac{\sigma_{12} I_s}{\hbar \omega_s} \right) + n_2 \left( \frac{\sigma_{13} I_p}{\hbar \omega_p} + \frac{\sigma_{12} I_s}{\hbar \omega_s} \right) + n_2 \left( \frac{\sigma_{21} I_s}{\hbar \omega_s} + \tau_{21}^{-1} \right) = 0$$

$$\left( \frac{\sigma_{13} I_p}{\hbar \omega_p} + \frac{\sigma_{12} I_s}{\hbar \omega_s} \right) = n_2 \left( \frac{\sigma_{13} I_p}{\hbar \omega_p} + \frac{\sigma_{12} I_s}{\hbar \omega_s} \right) + n_2 \left( \frac{\sigma_{21} I_s}{\hbar \omega_s} + \tau_{21}^{-1} \right)$$

$$\to n_2 = \frac{\frac{\sigma_{13} I_p}{\hbar \omega_p} + \frac{\sigma_{12} I_s}{\hbar \omega_s}}{\frac{\sigma_{13} I_p}{\hbar \omega_p} + \frac{(\sigma_{12} + \sigma_{21}) I_s}{\hbar \omega_s} + \tau_{21}^{-1}} = \frac{\frac{\hbar \omega_s \sigma_{13} I_p + \hbar \omega_p \sigma_{12} I_s}{\hbar^2 \omega_p \omega_s}}{\frac{\hbar \omega_s \sigma_{13} I_p + \hbar \omega_p (\sigma_{12} + \sigma_{21}) I_s}{\hbar^2 \omega_p \omega_s} + \tau_{21}^{-1}}$$

For $I_p \gg I_s$, which will be in this case, above equation can be approximated to:

$$n_2 = \frac{1}{1 + \frac{\hbar \omega_p \tau_{21}^{-1}}{\sigma_{13} I_p}} \quad (4.2\text{-}2)$$

Note that in addition to pumping the unstable $^4I_{11/2}$ level using $\lambda_p$ = 980 nm, it is possible to pump $Er^{3+}$ ions to the higher levels of $^4I_{13/2}$ Stark manifold directly using $\lambda_p$ = 1480 nm.

For $n_2 > 0.5$ amplification occurs. The next step is to calculate effective absorption coefficient $\alpha(\omega)$ in erbium doped garnets. This will include the absorption of undoped garnet, $\alpha_0(\omega)$, and gain generated by Er dopants. Based on the abovementioned assumption $n_3 \to 0$, gain can be expressed as [6]:

$$g = \sigma_e N_2 - \sigma_a N_1 = \sigma_{21} N_2 - \sigma_{12} N_1 \quad \text{where} \quad N_2 + N_1 = \rho_0$$

Here $\rho_0$ is the Erbium ion density, $N_1$ and $N_2$ are the population densities of Erbium ions in $^4I_{15/2}$ and $^4I_{13/2}$ energy states all in cm$^{-3}$ units. Following the McCumber relation, we can relate the Erbium's absorption to its emission cross section by:

$$\sigma_{21} = \sigma_{12} e^{\hbar \frac{\omega_{21} - \omega}{k_B T}}$$

$$\to g = N_2 \sigma_{12} e^{\hbar \frac{\omega_{21} - \omega}{k_B T}} - \rho_0 \sigma_{12} + N_2 \sigma_{12}$$

$$\left\{ n_2 = \frac{N_2}{\rho_0} \right\} \to g = -\sigma_{12} \rho_0 \left\{ 1 - n_2 \left( 1 + e^{\hbar \frac{\omega_{21} - \omega}{k_B T}} \right) \right\}$$

Here $n_2$ is the relative population densities of Erbium ions in $^4I_{13/2}$ energy state ranging from 0 to 1. Clearly, $\sigma_{12} \rho_0$, denoted as $\alpha_{Er}(\omega_{21})$ hereafter, is the maximum signal absorption in a host having Erbium dopant concentration $\rho_0$ when signal overlap factor $\Gamma_s = 1$, see equation 1.120 in [6]. To obtain the dispersive relation for gain, we replace $\alpha_{Er}(\omega_{21})$ with $\alpha_{Er}(\omega)$.

$$\to g_{Er}(\omega) = -\alpha_{Er}(\omega) \left\{ 1 - n_2 \left( 1 + e^{\hbar \frac{\omega_{21} - \omega}{k_B T}} \right) \right\}$$

We can then subtract the gain from the absorption of undoped garnet matrix $\alpha_0(\omega)$, balanced by factor $\rho'$ in due to Erbium concentration. This gives the total absorption coefficient $\alpha(\omega)$ in erbium doped garnets:

$$\alpha(\omega) = (1 - \rho') \alpha_0(\omega) + \alpha_{Er}(\omega) \left\{ 1 - n_2 \left[ 1 + \exp\left( \hbar \frac{\omega_{21} - \omega}{k_B T} \right) \right] \right\} \quad (4.2\text{-}3)$$

Here $\omega_{21}$ is a frequency of the resonance transition $2 \to 1$ at $\lambda_{21}=1531$nm. Expanding $\alpha=4\pi k/\lambda$ and $\omega=2\pi c/\lambda$ in above equation gives:

$$k(\lambda) = (1 - \rho') k_0(\lambda) + k_{Er}(\lambda) \left\{ 1 - n_2 \left[ 1 + \exp\left( \frac{2\pi c \hbar}{k_B T} \left( \frac{1}{\lambda_{21}} - \frac{1}{\lambda} \right) \right) \right] \right\} \quad (4.2\text{-}4)$$

Similarly, for the real part of the refractive index for Erbium doped garnet we get:

$$n(\lambda) = (1 - \rho') n_0(\lambda) + n_{Er}(\lambda) \quad (4.2\text{-}5)$$

Although Equation (4.2-4) is independent of cross sections due to McCumber relation, one must consider the range of wavelength in which this equation is deemed to be valid. A typical emission and absorption cross section spectra for Erbium has a FWHM of 40-60nm [18] around the resonance wavelength $\lambda_{res} = 1531$ nm caused by separation of sublevels within in $^4I_{15/2}$ and $^4I_{13/2}$ Stark manifolds. Furthermore, both emission and absorption cross sections approach 0 for $1450 > \lambda > 1630$ nm. Therefore 1450 < λ < 1630 nm is the widest possible wavelength range in which equation Equation (4.2-4) is deemed to be valid.

Values for $n_{Er}$ and $k_{Er}$ may be calculated in the following way. Here we assume that the contribution to the complex refractive index is purely due to the Erbium atoms. $Er_2O_3$ has a density of 8.64 g/cm$^3$ and a molar mass of 2×167.259 + 3×15.9994 = 382.51 g/mol. Density in terms of number of $Er_2O_3$ per unit volume = density × Avogadro's number /molar mass = 1.359×10$^{22}$ cm$^{-3}$. On the other hand, number of formula units per unit volume for $Gd_3Ga_5O_{12}$ was found to be 4.213×10$^{21}$ cm$^{-3}$ in previous chapters. Assuming that the variation in lattice constant due to Erbium dopants is negligible, $Er^{3+}$-ion concentration in $Gd_2Er_1Ga_5O_{12}$, is then 1×4.213×10$^{21}$ cm$^{-3}$. This corresponds to 0.5×4.213×10$^{21}$ = 2.106×10$^{21}$ cm$^{-3}$ $Er_2O_3$ compound in garnet. Therefore the $Er_2O_3$ in $Gd_2Er_1Ga_5O_{12}$ contributing factor is $\rho' = 2.106\times10^{21}/1.359\times10^{22} = 0.156$, thus $n_{Er} = \rho' n_{Er_2O_3}$ and $k_{Er} = \rho' k_{Er_2O_3}$. For $Gd_{3-x}Er_xGa_5O_{12}$ general expression for $\rho'$ in terms of Erbium concentration in garnet formula unit is given by $\rho' = 0.156x$, where x represents Erbium concentration in garnet formula unit

Complex refractive index for $Er_2O_3$ was determined experimentally in [18]:

$$\begin{aligned}(n_{Er_2O_3} &- ik_{Er_2O_3})^2 \\ &= 1 + \frac{2.66}{1-\left(\frac{150\ nm}{\lambda}\right)^2 + i3.61\times10^{-3}\times(\frac{150nm}{\lambda})} \\ &+ \frac{1.07\times10^{-4}}{1-\left(\frac{523\ nm}{\lambda}\right)^2 + i5.06\times10^{-3}\times(\frac{523nm}{\lambda})} \\ &+ \frac{7.24\times10^{-5}}{1-\left(\frac{528\ nm}{\lambda}\right)^2 + i6.81\times10^{-3}\times(\frac{528nm}{\lambda})} \\ &+ \frac{4.63\times10^{-5}}{1-\left(\frac{533\ nm}{\lambda}\right)^2 + i1.29\times10^{-2}\times(\frac{533nm}{\lambda})} \\ &+ \frac{7.62\times10^{-4}}{1-\left(\frac{1531\ nm}{\lambda}\right)^2 + i3.70\times10^{-2}\times(\frac{1531nm}{\lambda})}\end{aligned} \quad (4.2\text{-}6)$$

Thus the refractive index $n_{Er_2O_3}(\lambda) = \Re\sqrt{\varepsilon_{xx}}$ and extinction coefficient $k_{Er_2O_3}(\lambda) = -\Im\sqrt{\varepsilon_{xx}}$. And since for very transparent rare earth gallium garnets, GGG, $k_0\approx0$, we can then rewrite Equations (4.2-4) and (4.2-5) as:

$$k_{Er:GGG}(\lambda) = \rho' k_{Er_2O_3}(\lambda)\left\{1 - n_2\left[1 + exp\left(\frac{2\pi c\hbar}{k_B T}\left(\frac{1}{\lambda_{res}} - \frac{1}{\lambda}\right)\right)\right]\right\} \quad (4.2\text{-}7)$$

$$n_{Er:GGG}(\lambda) = (1-\rho')\, n_0(\lambda) + \rho' n_{Er_2O_3}(\lambda) \quad (4.2\text{-}8)$$

$$N_{Er:GGG}(\lambda) = n_{Er:GGG}(\lambda) - ik_{Er:GGG}(\lambda)$$
$$= [(1 - \rho') n_0(\lambda) + \rho' n_{Er_2O_3}(\lambda)] \qquad (4.2\text{-}9)$$
$$- i\rho' k_{Er_2O_3}(\lambda) \left\{ 1 - n_2 \left[ 1 + \exp\left(\frac{2\pi c\hbar}{k_B T}\left(\frac{1}{\lambda_{res}} - \frac{1}{\lambda}\right)\right) \right] \right\}$$

Once $n_{Er:GGG}(\lambda_{res})$ is known, thicknesses of Er:GGG layers must be determined to be an odd integer of a quarter resonance wavelength $\lambda_{res}/4n_{Er:GGG}(\lambda_{res})$ in Bragg reflectors and integer of a half resonance wavelength $\lambda_{res}/2n_{Er:GGG}(\lambda_{res})$ in microcavities.

### 4.2.1   GGG as Gain Media

The most fundamental design goals of the target AMPOC are as follow:

1. Maximum FR. The first criterion is achieved by maintaining highest Bismuth concentration in BIG layers, which also partially contributes to criterion 3. Undoped BIG parameters are identified in [19] which are used in all simulations. Therefore only Erbium doping was limited to GGG layers.
2. Minimum signal to noise ratio, thus minimum output noise power. This criterion is related to the value of spontaneous emission factor [6]: $n_{sp} = \frac{\eta n_2}{\eta n_2 - n_1}$ where $\eta = \frac{\sigma_e}{\sigma_a}$ and $n_1$ and $n_2$ are the relative population densities of $^4I_{15/2}$ and $^4I_{13/2}$ levels respectively. Minimum output noise power which is related to the maximum spontaneous emission factor is achieved at full medium inversion, $n_1 = 0$ and $n_2 = 1$ at high pumping power regime. This corresponds to the 3dB limit optical noise referred to as quantum limit. Therefore, for all simulation, a high value of 0.985 for inversion population is used, i.e. close to unity within realistic frame of expectations. In case one needs to identify commercially available lasers capable of delivering such intensity, Equations (4.2-2) may be solved for $I_p$ for a given $n_2$, $\sigma_{13}$ and $\tau_{21}$. As an indication, pump power needed to deliver such high inversion population in $Gd_{2.47}Er_{0.53}Ga_5O_{12}$ was found to be $P_p = 468$ mw@980nm assuming effective pump mode radius of $r_p = 3\times10^{-6}$, $\sigma_{13} = 1.7\times10^{-21}$ cm$^{-2}$ [20] and $\tau_{21} = 3$ ms, i.e. a reduction to what was reported in [5] for $Gd_{2.9}Er_{0.1}Ga_5O_{12}$ due to increase in Erbium concentration.
3. Minimum Thickness. Third criterion is related to optimization of the microcavity position and the configuration of the multilayer structure.

# 5. The Program

## 5.1   Description

Simulation program was developed in MATLAB to incorporate all the above mentioned theoretical concepts. Program was designed based on design goals such as modularity and parameterization, featuring:

1. Modules that model physical properties such as *n, k, $\varepsilon_{xx}$* and *$\varepsilon_{xy}$*
2. Modules that model mathematical relations such as Swanepoel formula and Višňovský's 4×4 matrix formalism
3. Parameterized input file in the form of spreadsheet containing configuration of the multilayer structure being examined where a layer within the target MOPC is represented by a row in the input file. This ensures decoupling of the input file/configuration from the MATLAB code. (simple software engineering best practices)

For magneto-optics material, a row contains dipole transition parameters, erbium doping concentration, inversion population parameter and thickness. For transparent layers and substrate, each row contains Sellmeier parameters, erbium doping concentration, inversion population parameter and thickness. Separation of physical and computational models provides a strong versatility in searching and identifying multilayer structure with desired properties.

Program treats each row independently from other rows using only the relevant parameters specified for a given layer. This gives a great degree of freedom to model very realistic scenarios where, for example, physical property of a given material such as GGG may or may not be identical to other GGG layer. An example would be a scenario where GGG layers closer to the substrate have higher erbium doping in comparison to those on top.

For a given wavelength, or wavelength range, the program transverse the multilayer structure producing output files containing transmission, FR and ellipticity spectra.

Development of the program was carried out by me solely. Therefore, the copy right will rest with me alone and as such code will not be listed here.

# 6. Results and Analysis

## 6.1 Calibration

At early stages of the project, simulation results were examined against the available experimental data for undoped BIG film, see Figure 7(a), and BIG/SGG MOPC, see Figure 7(b), reported in [21]. Close agreements between the simulated and experimental data for FR and transmission spectra in both cases, indicate the merit of theoretical analysis.

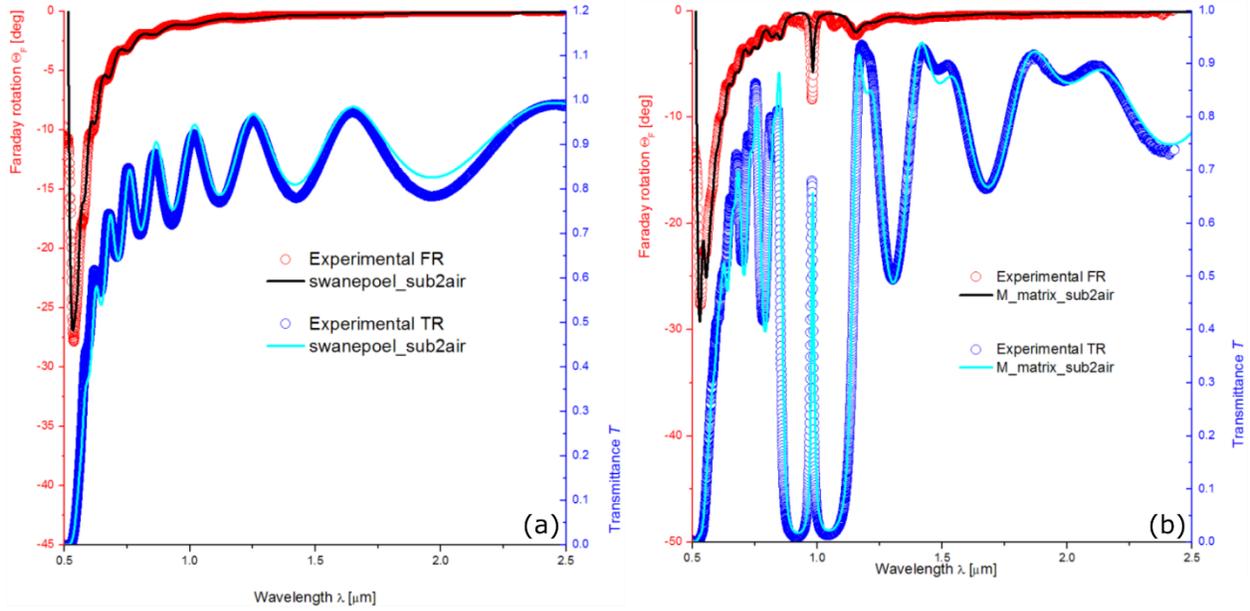

**Figure 7: (a)** Simulated and experimental transmission and Faraday rotation spectra for 1μm $Bi_3Fe_5O_{12}$ epitaxially grown on Ca,Mg,Zr:$Gd_3Ga_5O_{12}$ substrate. **(b)** Simulated and transmission and Faraday rotation spectra for $[BIG/SGG]^5$ $BIG^2$ $[SGG/BIG]^5$ epitaxially grown on Ca,Mg,Zr:$Gd_3Ga_5O_{12}$ substrate.

## 6.2 AMOPC

### 6.2.1 Effect of Erbium Concentration

A multilayer structure with a configuration $[B/G]^{10}$ B $G^5$ B $G^8$ B $G^3$ $[B/G]^{10}$ B was initially examined under 0.985 inversion population with different Erbium concentration to obtain the target MO effects. Here the substrate is located to the right end of structure. In this context, $B^j$ stands for BIG layer having thickness of $j \times \lambda_{res}/4n_{BIG}(\lambda_{res})$, $G^j$ stands for GGG layer having thickness of $j \times \lambda_{res}/4n_{GGG}(\lambda_{res})$ and $[B/G]^m$ represents 'm' B/G pairs. It was found that Erbium concentration, in garnet formula unit, of x = 0.53 in GGG layers of the above mentioned AMOPC, produces FR $40° < FR < 45°$ and transmittance of $0.5 < T < 0.85$ for $1530.8 < \lambda < 1531.2$nm, see Figure 8(a), with minimum ellipticity of $6°$, see Figure 8(b), at $\lambda_{res} = 1531$ nm where B = 154.43 nm and G = 197.3243 nm.

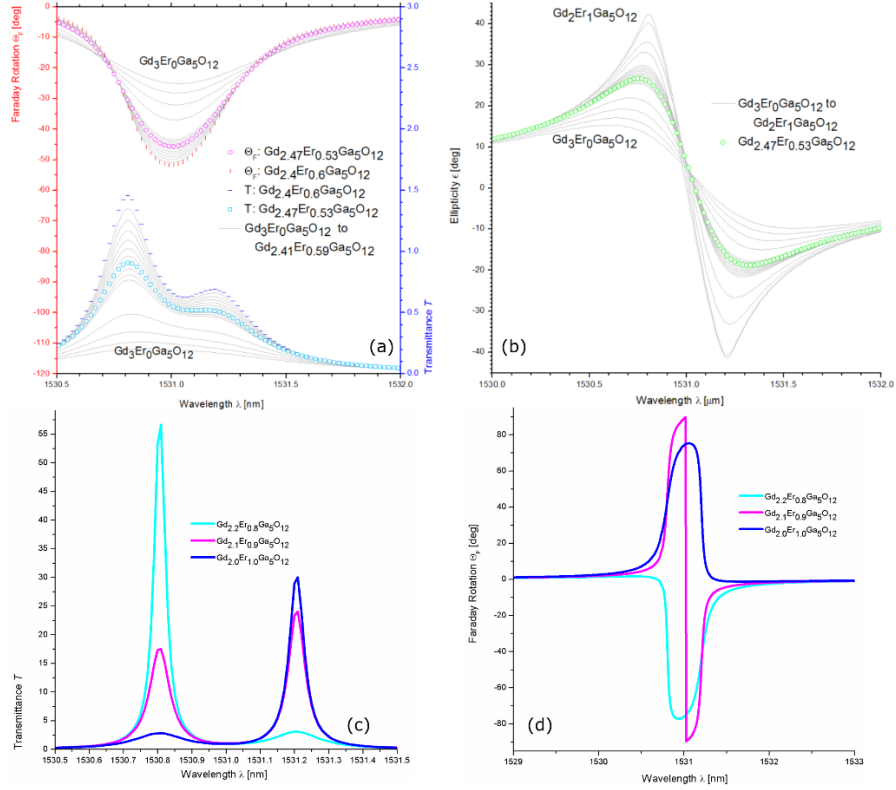

**Figure 8:** (a) Transmission and Faraday rotation spectra for [B/G]$^{10}$ B G$^5$ B G$^8$ B G$^3$ [B/G]$^{10}$ B with various Erbium concentration. (b) Ellipticity spectra for [B/G]$^{10}$ B G$^5$ B G$^8$ B G$^3$ [B/G]$^{10}$ B with various Erbium concentration. (c) Transmission spectra for Erbium concentration 0.8, 0.9 and 1 garnet formula unit in MOPC with [B/G]$^{10}$ B G$^5$ B G$^8$ B G$^3$ [B/G]$^{10}$ B configuration. (d) Faraday rotation spectra for Erbium concentration 0.8, 0.9 and 1 garnet formula unit in MOPC with [B/G]$^{10}$ B G$^5$ B G$^8$ B G$^3$ [B/G]$^{10}$ B configuration. Note the edge state at $\lambda_{res}$ = 1531 nm for 0.9 Erbium concentration.

## 6.2.2 Microcavity Position

FR and transmittance of the abovementioned AMOPC was further enhanced by examining different microcavity position. An interesting effect was observed when microcavity position was shifted away from the substrate. Moving from **[B/G]$^{10}$ B G$^5$ B G$^8$ B G$^3$ [B/G]$^{10}$ B** to **[B/G]$^9$ B G$^5$ B G$^8$ B G$^3$ [B/G]$^{11}$ B** produces highest transmittance and FR, thus optimum, with almost no significant change to ellipticity, see Figure 9(a) and (b). However, further shift away from the substrate by one B/G pair, i.e. **[B/G]$^8$ B G$^5$ B G$^8$ B G$^3$ [B/G]$^{12}$** greatly reduces the MO effect. This may indicate existence of a maxima/minima profile associated to MO properties within the multilayer structure, whereby positioning the microcavity in a maxima would result in an optimum condition.

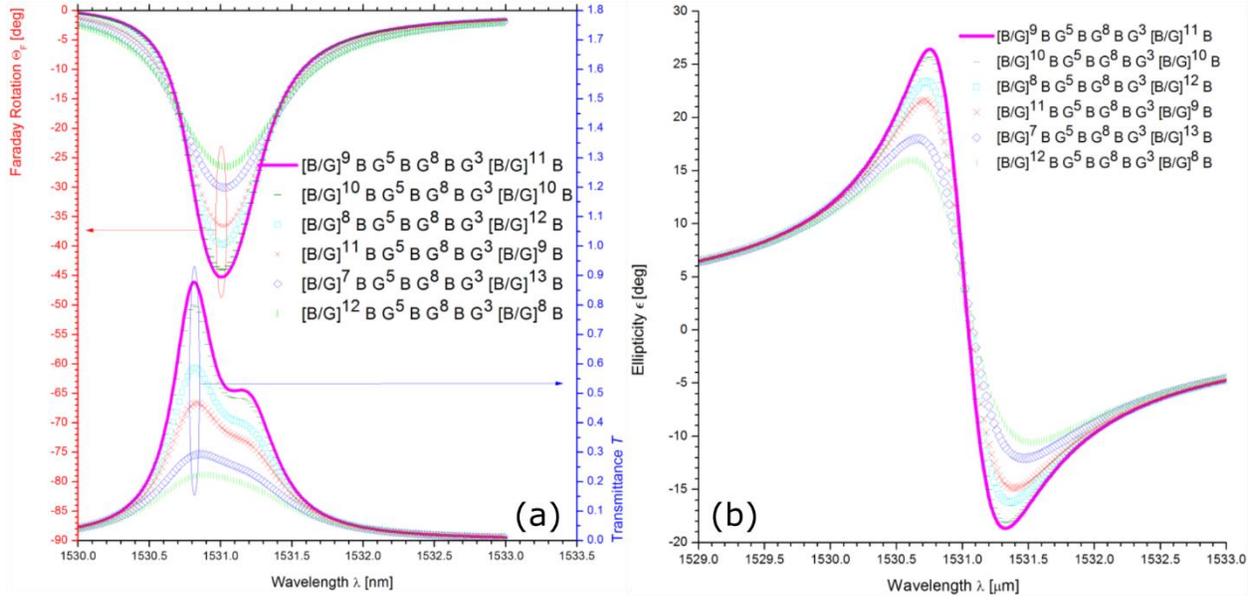

**Figure 9: (a) Transmission and Faraday rotation spectra for various microcavity positions within the MOPC having Erbium concentration x=0.53. (b) Ellipticity spectra for various microcavity positions within the MOPC having Erbium concentration x=0.53.**

### 6.2.3 Optimum Configuration

For the optimum configuration above, in the flat region of transmission spectra $1351.0 < \lambda < 1351.3$ nm, FR and transmittance ranged $-44° < FR < -46°$ and $0.5 < T < 0.65$ respectively. With $-46°$ FR and 0.9 transmittance at their peak positions, see Figure 10(a), for a AMPOC with a total thickness of 10.8 µm, specific FR of $-4.26°/\mu m$ and MO figure of merit $Q = \Theta_F/\ln(1/T) = 359°$ at $\lambda = 1351$ nm is obtained. This is an enhancement of $-1.16°/\mu m$ compared to a single layer $Bi_3Fe_5O_{12}$ film and an enhancement of $-0.32°/\mu m$ and $83°$, respectively, to what was reported in [22] at half the erbium concentration and 0.6µm less in thickness.

For the optimum AMOPC above, transmittance at $\lambda_{pump} = 980$ nm and 1480 nm were limited to 0.5 and 0 respectively by the Bragg reflectors, see Figure 10(b). Setting the resonance wavelength of the MOPC to $\lambda_{res} = 1531$ nm by design, may or may not guarantee a high enough transmittance for pump power at $\lambda_{pump} = 980$ nm and 1480 nm at normal incident, to create a target inversion population of 0.985 in all layers. Being a theoretical device, it is assumed that pump wave at lateral incident, i.e. propagation in the direction parallel to the layers' surfaces, see Figure 5, would achieve the target population inversion. Variation of temperature in the range of 100K-500k, due to inclusion of McCumber relation, showed no significant change to MO effects of AMOPC, see Figure 10(c)&(d).

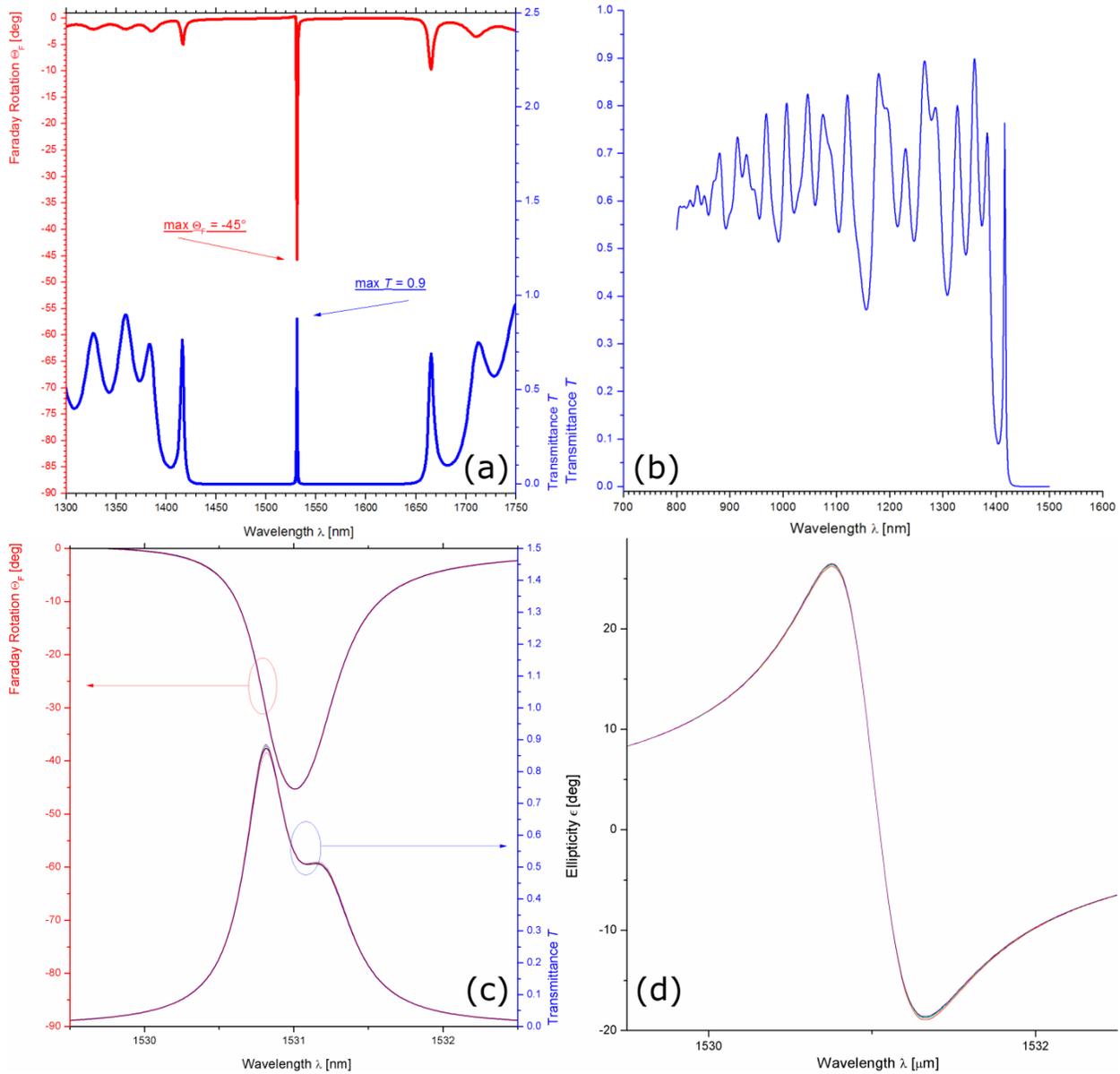

**Figure 10:** Results for [B/G]$^9$ B G$^5$ B G$^8$ B G$^3$ [B/G]$^{11}$ B. **(a)** Transmission and Faraday rotation spectra for Erbium concentration x = 0.53 under high pumping regime associated to inversion population $n_2$ = 0.985. **(b)** Transmission spectrum for 800 < λ < 1500 nm used for determination of transmittance at pump wavelengths $\lambda_{pump}$ = 980 nm and 1480 nm. **(c)** Transmission and Faraday rotation spectra for Erbium concentration x = 0.53 under high pumping regime associated to inversion population $n_2$ = 0.985 at temperature range 100 to 500K. **(d)** Ellipticity spectra for Erbium concentration x = 0.53 under high pumping regime associated to inversion population $n_2$ = 0.985 at temperature range 100 to 500K.

## 6.2.4 What Was Published

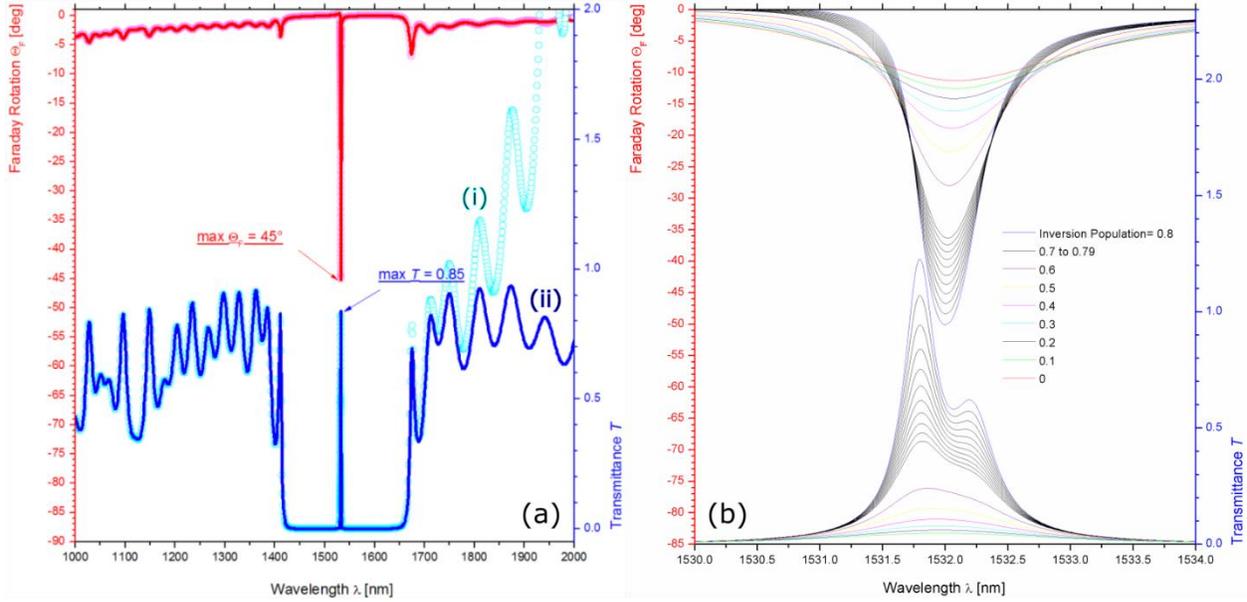

**Figure 11: (a) Transmission and Faraday rotation spectra for AMOPC reported in [22] recalculated with (i) full MacCumber relation included – Cyan ○ and (ii) exponent in McCumber relation replaced with 1 – blue line. (b) Transmission and Faraday rotation spectra for AMOPC reported in [22] recalculated for various values for inversion population with McCumber relation included.**

The author in [22] claims the transmittance in figure 1[22], was produced by a program that implemented equations (1) and (2) [22], with full McCumber relation that includes the $\exp\left(\frac{2\pi c\hbar}{k_B T}\left(\frac{1}{\lambda_{res}} - \frac{1}{\lambda}\right)\right)$. However, when I reproduced the same, the result was Figure 11(a)-(i). I also reproduced another spectrum with $\exp\left(\frac{2\pi c\hbar}{k_B T}\left(\frac{1}{\lambda_{res}} - \frac{1}{\lambda}\right)\right) = 1$, see Figure 11(a)-(ii), which is exactly the same spectrum reported in [22]. The explanation is simple. Almost 6 months into my project work and shortly after submitting a joint patent application in late June 2010, my supervisor at the time, (the author in [22]), requested that I install my program on his workstation in his office, (KTH/Kista), instructing him how to prepare an input configuration file and how to run the MATLAB program…etc. However, in the version I installed on his workstation, I had replaced the exponent to unity, that is $\exp\left(\frac{2\pi c\hbar}{k_B T}\left(\frac{1}{\lambda_{res}} - \frac{1}{\lambda}\right)\right) = 1$ which implied λ = λ$_{res}$ for all wavelengths. Given his limited to no programming skills, the author in [22] had no way of knowing this. But I did emphasize that my program was a work in progress. Although the results reported in [22] are valid for λ = λ$_{res}$, and I am not disputing that, it shows that the author in [22] could not have produced the results with any program other than the one I installed on his workstation in late June 2010. In fact, prior to the publication of [22] on August 2010, I had absolutely no idea that my program had been utilized without my permission. And being mentioned in the acknowledgment doesn't change anything. Perhaps the author in [22] assumed that by altering the input configuration file (which is nothing but a spreadsheet) and proposing an alternate layered structure, he reserved the rights to utilize the MATLAB code I developed. But no amount of modifications to the input file changes the MATLAB code nor does it justify the unauthorized use of my program. Furthermore, the author associated my work with some fund provided by Swedish Research Council that I was neither part of nor aware of. Consequently, our relationship was severed, the patent application was abandoned, and I had to write my thesis [23] unsupervised.

The point I am trying to make here, is the amount of questionable practices I have witnessed (if not subjected to) by academics in the past decade or so. Certainly, unethical subjugations and exploitations by some high-ranking academics has become systemic as I experienced and was subjected to such malpractices in yet another institution in Australia for years that followed.

# 7. Conclusion and Future Work

Modelling Erbium doped MOPCs shows promising results. Overcoming the long-standing encumbrance of large insertion loos associated to thick MOPCs is shown to be possible in a theoretical sense. Two interesting effects were observed which remains unexplained. Under the high pump power regime, with erbium concentration of 1 garnet formula unit, the shoulder at $\lambda = 1531.2$ nm becomes a fully developed peak surpassing the original peak at $\lambda = 1530.8$nm, Figure 8(c). At this concentration FR changes sign, and with 0.9 Erbium concentration produces an edge state at $\lambda = 1531$ nm, Figure 8(d). These transitional effects are also reproducible by maintaining high erbium dopant concentration with varying pump power. The main reason for discussing it under this heading is to perhaps instigate a future work for those interested. Clearly if such an effect is physically realizable, it can be deployed in modulators.

As for the merit of our simulated results, I must mention that Erbium's emission and absorption cross sections vary depending on the host. These variations are due to the Stark sublevels' splitting caused by the lattice, in this case Er:GGG, thus every host has unique finger print on emission and absorption cross sections. Furthermore, comparing the Erbium concentration in $Er_2O_3$ to Er:GGG, one can conclude that even at high dopant concentration, Erbium atoms in GGG are located further apart from each other, resulting in a less likelihood of intermixing of energy levels with those of the neighboring dopant atoms. Therefore, energy levels of $Er^{3+}$-ions in GGG are perceived to be more discrete compared to those in $Er_2O_3$, which also has a direct impact on the emission and absorption cross sections. Although, in absence of cross sections for $Er^{3+}$-ions in GGG, inclusion of McCumber relation in conjunction with $Er_2O_3$ data may provide us with a reasonable approximation, an accurate model is only possible when emission and the absorption cross sections in GGG are measured and modelled using Lorentzian fit.

**Disclosure**
The author (Amir Djalalian-Assl) received no remuneration of any sort, from any government or institution, during his master program 2008-2010, including the duration of the project work reported here. Associating any part of the original research in this report, previously published or otherwise, to any funding scheme is considered misrepresentations of the author's original work.